# An automatic mixing speech enhancement system for multi-track audio

Xiaojing Liu[1], Angeliki Mourgela[1], Hongwei Ai[2] and Joshua D. Reiss[1]
[1]Centre for Digital Music, Queen Mary University of London, UK
[2]Independent Researcher
Correspondence should be addressed to Xiaojing Liu(xiaojing.liu@qmul.ac.uk)

## Abstract

We propose a speech enhancement system for multitrack audio. The system will minimize auditory masking while allowing one to hear multiple simultaneous speakers. The system can be used in multiple communication scenarios e.g., teleconferencing, in-voice gaming, and live streaming. The ITU-R BS.1387 Perceptual Evaluation of Audio Quality (PEAQ) model is used to evaluate the amount of masking in the audio signals. Different audio effects e.g., level balance, equalization, dynamic range compression, and spatialization are applied via an iterative Harmony searching algorithm that aims to minimize the masking. In the subjective listening test, the designed system can compete with mixes by professional sound engineers and outperforms mixes by existing auto-mixing systems.

## 1. Introduction

The cocktail party effect [1] refers to the phenomenon that humans can focus on a specific sound or conversation while filtering out other sounds in a noisy environment, such as a restaurant or a reception. Studies on auditory attention and selective hearing have deepened the understanding of the cocktail party effect and revealed how the brain distinguishes sound sources by analyzing the spatial localization and frequency characteristics of sounds [1]. Advances in neuroscience research have allowed scientists to study the brain's processing in multiple sound source environments in greater detail. The researchers [1] found that the auditory cortex exhibits a high degree of dissociation and adaptability when processing information from multiple sound sources, which helps explain human auditory performance in scenes such as cocktail parties.

Researchers explored how the cocktail party effect could be applied to problems such as source separation [2] and voice enhancement [3]. Some voice enhancement aimed to extract a target speaker in a multi speaker environment, and reduce the unwanted sources, environmental noise, or reverb [4]. However, the majority of voice enhancement work focused on extracting a single target voice and ignoring other tracks that might have included important information.

The work of [5] also indicated the problem of perceiving sounds in information losing: Informational masking was a multifaceted phenomenon resulting from various stages of processing beyond the auditory periphery. It was closely tied to perceptual grouping, source segregation, attention, memory, and broader cognitive processing abilities, highlighting the intricate interplay between auditory perception and higher-level cognitive functions.

G. Kidd Jr et al. [6] found that when the target sequence was spatially separated from the masker, it resulted in a significant enhancement of the segregation between the target and the masker. This spatial separation facilitated the listener's ability to concentrate attention on the target, leading to a considerable reduction in informational masking. Rothbucher et al. [7] also researched a teleconferencing system combining HRTF with VoIP (Voice over IP) that could perform online sound localization, source separation, speaker detection, and channel allocation. The work [7] above was related to the research of teleconferencing (multiple speaker scenarios) and considered to improve the user experience through audio enhancement. However, the work of Rothbucher et al considered the microphone arrangement but did not quantify or consider the masking effect.

The work of [8] exploited time-frequency masking from multiple microphones to distinguish different sound sources, thereby improving the accuracy and effectiveness of speech enhancement. However, the approach required extensive computing resources for training and inference. Furthermore, spatial filtering would result in masking for other tracks.

In [9], an automatic mixing system for teleconferencing was proposed. It aimed to improve audio quality by, among other things, use of a graphic equalizer to achieve equal average perceptual loudness in each frequency band. However, this approach might result in even more frequency masking.

In this paper, we propose a lightweight system using iterative optimization for multi-track audio. The system will minimize the masking so that other tracks can be heard when more than one source is active. The PEAQ (Perceptual Evaluation of Audio Quality) [10] model will be used to evaluate auditory masking in the system. The masking value serves as an input vector for optimising the parameters of applied audio effects including equalization, dynamic range compression, and spatialization. The range of parameters of audio effects will be set according to best practices in the audio engineering field. The system will iterate each parameter until the masking model is reduced to zero or the system reaches some maximum number of iterations.

Objective tests first give insight into the algorithm's performance, while subjective tests involve comparing the proposed work against mixes created by professional sound engineers and existing auto-mixing systems [9].

## 2. The proposed framework

The proposed system is shown in Figure 1. An improved PEAQ model will be used to measure masking in the system. The masking value is then used to optimize the parameters of applied audio effects. The system will process the input audio tracks through different sound effect blocks and process them separately. The harmony searching algorithm is used for iterative optimization.

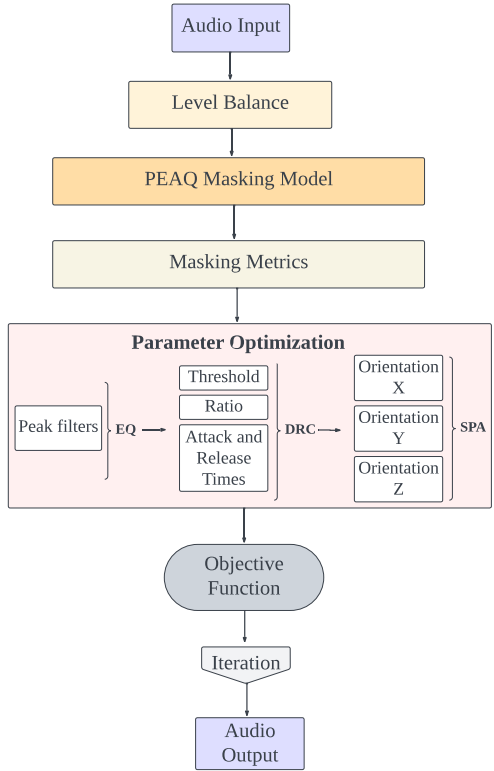

Figure 1: The workflow of proposed system

### 2.1 Level Balance

LUFS, or Loudness Units Full-Scale, serves as a standardized measure for evaluating sound loudness, taking into account both human perception and electrical signal strength [11]. A the EBU R 128 guidelines, the recommended loudness level for radio programs is set at -23 LUFS [11]. The method outlined in reference [12] is employed to calculate and compensate for the loudness of each track, aligning it with the LUFS standard loudness level. This process ensures a consistent and optimal loudness experience across various tracks, adhering to industry standards for broadcasting.

### 2.2 PEAQ Model

PEAQ is used to objectively measure audio quality [10], and elements from the PEAQ algorithm can be employed to make informed design choices for audio coders. The model simulates aspects of human hearing to estimate thresholds for masking of audio signals. Figure 2 describes the workflow of the PEAQ model in the system we implemented.

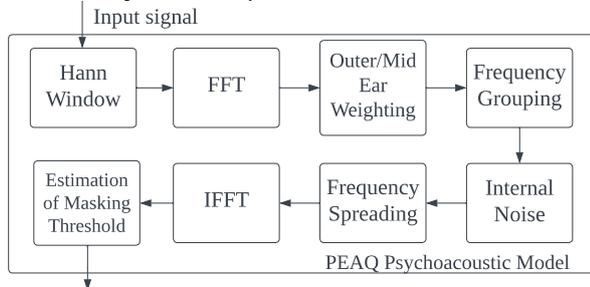

Figure 2: The workflow of PEAQ psychoacoustic model

The estimation of masking thresholds is as follows. First, the input signal is transformed to the frequency domain using the Fast Fourier Transform (FFT) with a Hann window. Since sound pressure level affects perceptual quality, the spectrum energy loudness needs to be corrected by a weighting function to simulate the human ear's sensitivity curve in the Outer/Mid Ear function. In the critical band decomposition, it involves taking a frame of frequency domain samples and grouping them into the frequency bands. The Internal Noise part simulates the inner ear noise produced by the blood flow in the human ear [10]. Finally, frequency spreading simulates the smearing effect of wide auditory filters.

The masking threshold's energy distribution can be determined by using the masking curve, and we can determine the masking properties and energy of each critical band. Following the work of X. Hu et al [13]. The masking threshold (energy) for each critical band is obtained by:

$$E_{mask(dB)}(cb) = E_{f(dB)}(cb) - m_{dB}(cb) \quad (1)$$

$$m_{dB}(z) = \begin{cases} 3 & z \leq z_L + 12 \\ 0.25(z - z_L) & z > z_L + 12 \end{cases},$$

where $z$ is the central frequency of each band on bark scale (In the basic version of the PEAQ there are defined as 109 numbers [10]), and $z_L$ is equal to 0.8594.

All above steps are pertaining to processing a single frame, while the time spreading part works on multiple frames. In the masking threshold, the energy in each critical band, $Es(cb)$ is frequency spreading function which computed from [10] and the threshold in each critical band, $E_{mask(dB)}(cb)$ are calculated as described in Equation (1). Thus, the final masker-to-signal ratio (MSR) in each critical band (cb) is defined as

$$\text{MSR}(cb) = 10\log_{10}\left(\frac{E_{mask(dB)}(cb)}{E_s(cb)}\right) \quad (2)$$

### 2.3 Estimation of Masking Metrics

To estimate how much the masking track masked by other tracks in the whole mixing extend, we consider Zhen Ma et al [14]'s work: $T'_n(cb)$ is the masking threshold of track n caused by the sum of accompanying tracks, and the $E_{sf,n}(cb)$ is the energy of track n in each critical band computed as described in the formula (2) with different notation. The final masker-to-signal ratio (MSR) in each critical band is defined as

$$\text{MSR}_n(cb) = 10\log_{10}\left(\frac{T'_n(cb)}{E_{sf,n}(cb)}\right) \quad (3)$$

In the teleconferencing scenario, the $N$ is the total number of all the voice tracks, $T_{\max}$ is the range maximum amount of masking distance between T(cb) and $E_{sf,n}(cb)$ for each critical band, which is set to 20 dB.

$$M_n = \sum_{sb \subset E_{sf,n} < T'_n(cb)} \frac{\text{MSR}_n(cb)}{T_{\max}} \quad (4)$$

### 2.4 Objective Function

The minimization work of multitrack masking should connect with the parameter through system calculation with masking vector valued objective function. In [15], the aim is to build the unmasking model in the music production and each track's parameter should be computed until the error of the objective function the masking metric is reduced to zero or the system reaches the maximum iterations' number. A few parameters are introduced into the optimization algorithm before defining the objective functions. A is the total number of tracks, $\mathbf{x}_C$ is the function of all tracks' parameter control. The value of masking metrics is given by $M_i(\mathbf{x}_C)$ which is the masking value for $i$ tracks comparing with other sum tracks. The total number of masking is $M_T(\mathbf{x}_C)$ which consist of the sum of $M_i^2(\mathbf{x}_C)$. for i = 1 to A.

$$M_T(\mathbf{x}_C) = \sum_{i=1}^{A} M_i^2(\mathbf{x}_C) \quad (5)$$

The objective of the algorithm is to minimize the sum of the masking across tracks and so can be used as the first part of the objective function. The second objective is that the masking is balanced which means no difference between masking levels and a maximum masking difference is given by:

$$M_d(\mathbf{x}_C) = \max(\|M_i(\mathbf{x}_C) - M_j(\mathbf{x}_C)\|) \quad (6)$$
$$\text{for } i = 1, \dots, n, j = 1, \dots, n, i \neq j$$

From the parameter changes, the value of $\mathbf{x}_C$ wil affect the masking of that track itself but also masking of all other tracks.

$$\mathbf{x}_C = \max_{\mathbf{x}_C} M_T(\mathbf{x}_C) + M_d(\mathbf{x}_C) \quad (7)$$

### 2.5 Iteration

In the iteration section, we choose the Harmony Search [16] iterative algorithm for our system. The basic idea of the harmony search algorithm is to gradually improve the existing solution by continuously adjusting and combining the candidate solutions until the optimal solution is found. The process of running the algorithm is similar to the harmonization process in a band performance. In that, each musician (or harmony member) represents a variable of the optimization problem, and their performance level aligns with the quality of the solution. By continuously coordinating and adjusting the performance of each musician, one hopes to find an optimal performance combination.

In our system, the harmony search algorithm will randomly select different parameters with different effects in Equalization (EQ), Dynamic Range Compression (DRC), and Spatial Audio (SPA) through Web audio API [17]. In the EQ stage, every input signal undergoes gain modification via second order IIR filters within a filter bank that encompasses 8 frequency bands. The center frequencies of the equalizers are set at 60, 100, 200, 400, 800, 1600, 2500, and 7500Hz. The SPA employs Cartesian coordinates (X, Y, Z axis) for source positioning, using vectors for location and a 3D directional cone for orientation. To set the parameter range, professional practitioners in the field of audio engineering were consulted and tested, the parameter range of each audio effect is shown in Table 1. The masking value (after the PEAQ model and masking metrics) is used to measure the changes in the masking value in the objective function and the parameters are recorded for each iteration. Eventually, the system will retain the parameter value with the smallest masking value.

Table 1. The value range of EQ, DRC, and SPA parameters

| Audio Effects Process | Min Value | Max Value |
|---|---|---|
| EQ Gain Bands 1-8 | -15 dB | + 15 dB |
| DRC Ratio | 1 | 5 |
| DRC Threshold | -15 dB | 0 dB |
| DRC Attack | 0.01 secs | 0.5 secs |
| DRC Release | 0.05 secs | 1 secs |
| SPA XYZ Axis | -3 | 3 |

## 3. Experiment

To simulate multiple speaker scenes, the audio stimulus was collected from the LibriSpeech dataset [18] or extracted from multiple-speakers video in YouTube. Each scenario will be added, re-design, and clip the soundtracks. The stimuli used in the subjective listening tests contained three scenarios (Teleconferencing, Gaming, and Live streaming). Each scenario has a distinct number of tracks: 3 tracks for teleconferencing, 4 tracks for gaming, and 6 tracks for live streaming. Those files are 48000 Sample Rate, single-channel and 16 Bitrate.

### 3.1 Objective Analysis

Three objective tests were conducted to analyze the results concerning loudness, long-term average spectrum, and spatial positions for various tracks. Due to page limitations, we illustrate the analysis using the stimulus from the teleconferencing scenario as an example. Table 2 presents the loudness levels of each track before and after automatic mixing. It adeptly balanced the loudness in accordance with the LUFS standard, ensuring a harmonious audio output. The loudness levels are not equal since the EQ and DRC parameters are fine-tuned based on PEAQ's analysis of energy on each track, resulting in slightly different characteristics.

Table 2. *The LUFS comparison of the voice tracks before and after level balancing*

| File Name | Loudness Before | Loudness After |
|---|---|---|
| Total Track | -12.172 | -14.940 |
| Track1 | -27.279 | -19.882 |
| Track2 | -12.655 | -21.751 |
| Track3 | -44.064 | -20.343 |

Figure 3 (a) illustrates that the tracks 'spatial locations' would become separated from concentration. Figure 3 (b) shows that the 'unmix' figure indicates that Track 2 might dominate the frequency range of 150Hz to 10,000Hz, potentially overshadowing other sounds. This dominance could lead to Track 2 becoming the primary focus, with other sounds being masked or subdued. In the 'after mix' Figure, the frequency range in the mixed system is well-balanced between the left and right channels.

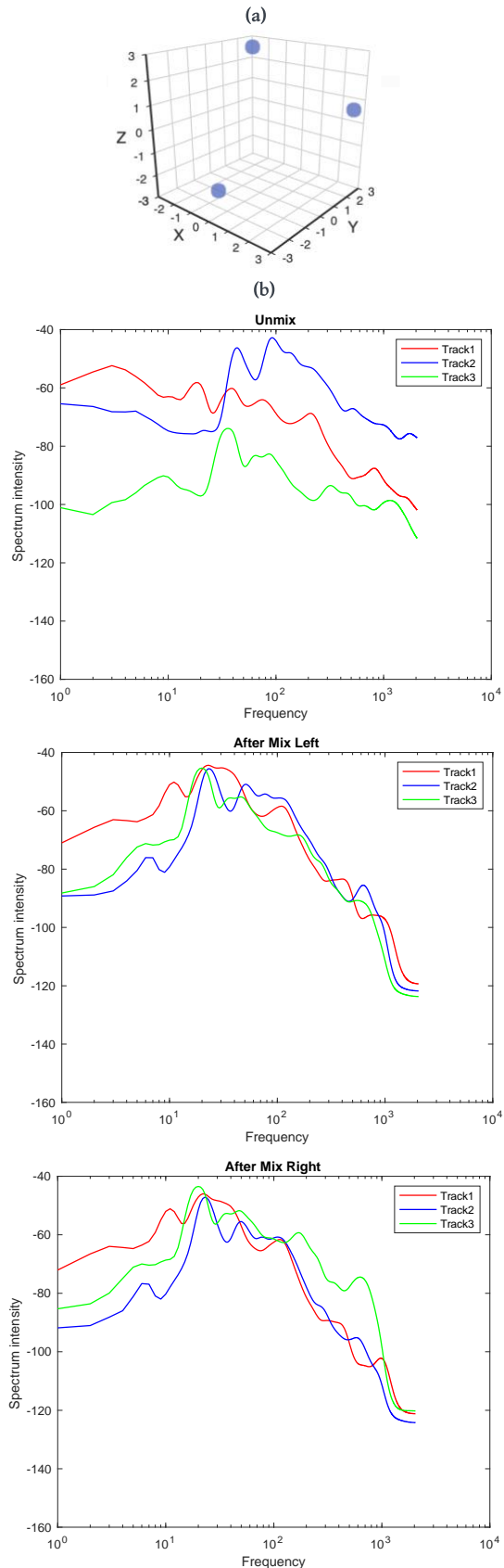

*Figure 3: The objective test results*

### 3.2 Subjective Evaluation

In the subjective listening test, we used the Go Listen platform [19] to conduct a blind comparison test. The materials of the subjective listening test included current automatic mixing system's output, the existing automatic mixing system's output from [9], manual mix output by an expert audio engineer, one from unmixed content, and two anchor versions of unmixed content (3.5khz and 7khz low pass filter).

A total of 16 participants took part in the test and were instructed to conduct the evaluation in a quiet environment. During the test, participants were required to evaluate audio clarity: whether they could recognize the track number easily in different scenes. Results are presented in Figure 4. The Auto mix has consistent ratings around 75 and outperforms the unmix, anchors, and existing automatic system in each scenario. In the teleconferencing scenario, the manual mix gets the highest average score. However, the average score of the manual mix is below the Auto mix in both gaming and live-streaming scenarios. A potential reason for this is the added complexity when attempting to manually process a larger number of tracks.

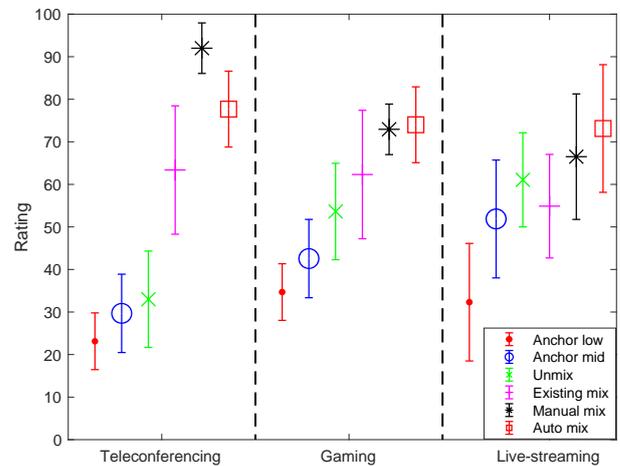

*Figure 4: The multi-stimulus test results, with 95% confidence intervals.*

## 4. Conclusions

Our study has provided valuable insights into a lightweight automatic multi speaker mixing system. The system aims to determine the optimal parameters for a targeted audio segment, designed for preprocessing in an audio scene. The results demonstrate that our system can compete with existing automix work and a manual mix in multiple speaker scenarios. However, for real-time implementation, it is crucial to focus on enhancing the algorithm's robustness and formulate effective strategies for adapting parameter adjustments to dynamic changes in the current audio environment.